\documentclass{article}

\usepackage[english]{babel}

\usepackage[letterpaper,top=2cm,bottom=2cm,left=3cm,right=3cm,marginparwidth=1.75cm]{geometry}

\usepackage{amsfonts}
\usepackage{amsmath,amssymb}
\usepackage{graphicx}
\usepackage[colorlinks=true, allcolors=blue]{hyperref}
\usepackage{multirow}

\begin{document} 

\title{A two-stage approach for table extraction in
invoices}

\author{ Thomas Saout \\
Univ Angers, LERIA, 
\\F-49000 Angers, France\\
thomas.saout@etud.univ-angers.fr
\and
Frédéric Lardeux\\
Univ Angers, LERIA, 
\\F-49000 Angers, France\\
frederic.lardeux@univ-angers.fr
\and
Frédéric Saubion \\
Univ Angers, LERIA, 
\\F-49000 Angers, France\\
frederic.saubion@univ-angers.fr}

\maketitle

\begin{abstract}
The automated analysis of administrative documents is an important field in document recognition that is studied for decades. Invoices are key documents among these huge amounts of documents available in companies and public services. Invoices  contain most of the time data that are presented in tables that should be clearly identified to extract suitable information. In this paper, we propose an approach that combines an image processing based estimation of the shape of the tables with a graph-based representation of the document, which is used to identify complex tables precisely. We propose an experimental evaluation using a real case application.
\end{abstract}


\section{Introduction}

The automated analysis of administrative documents is an important field in document recognition that is studied for decades \cite{BelaidDHB11}. Among these huge amounts of documents that are available in companies and public services, invoices are key documents. Their automated processing is a complex task \cite{DAndecyHR18} and has led to the development of commercial systems developed by companies such as ITESOFT or ABBYY.  

Invoices generally require complex administrative procedures, which involve different departments (e.g., accounting department, logistics, supply chain...). The invoices have to be processed through specific workflows \cite{Hollingsworth1994}. From the document point of view, the full processing of invoices includes their digitalization using Optical Character Recognition (OCR) \cite{Smith07} and their processing to achieve information extraction, which aims at finding identifiers and their types, amounts, dates \cite{HaH22,HamdiCJCD21}. This global process requires handling some specific characteristics of the considered documents \cite{DAndecyHR18}:
\begin{itemize}
\item 
handling the variability of layouts,
\item 
training and quickly adapt to new contexts,
\item 
minimizing the end-user task.
\end{itemize}

\noindent
{\bf Context}

Many solutions have been proposed to manage information from scanned invoices and most of them are based on machine learning techniques (e.g. classification). Recent research is still active on this problem \cite{HaH22,Patel2020}. The first problem was certainly to identify invoices \cite{KoppenWN96} and hence, models have been proposed to ease their processing \cite{CesariniFMSS97}. Once invoices have been correctly scanned and identified, a remaining crucial question is "how to extract relevant information from these invoices ?". Labeling techniques can be applied using rules \cite{DengelK02}. Addressing this named entity recognition (NER) task has been recently handled using neural networks \cite{LampleBSKD16,LiSHL22}. Since invoices contain text sequences that are mostly different from natural languages corpus, specific information extraction methods have been proposed to take into account the specific structures in these documents. For instance, in  \cite{DAndecyHR18}, a star graph is used to consider the neighborhood of a token (a token is an elementary, semantically coherent, element of the document). The specific structures of invoices lead to consider the geographical organization of the document and graph-based models are thus relevant \cite{ShafaitS10}. Examining more precisely invoices leads to consider that most of them include tables as a main structural character. 
Hence, table detection within invoices appears as an important processing task  \cite{ShafaitS10}. Table processing is indeed an old challenge \cite{EmbleyHLN06,ZanibbiBC04} and, as quoted in \cite{JhaN08} it includes different tasks : detection, extraction, interpretation and understanding. Here, we focus on detection (detecting the presence of a tabular structure in a document) and extraction (providing the data in a detected table in a more readable format). These challenges are still active \cite{GaoHDMYFKL19}. Recent work\cite{KashinathJAAS22} proposes an approach to detect the general frame of a table and to extract its content. Targeting wide classes of documents, many recent works often use neural networks \cite{KhanKSS19,GilaniQMS17,Riba0GFT019,LeePKC22} to recognize table structures in documents by means of large training sets. Focusing on more specific tables, their characteristics are also intended to help these tasks, such as headers \cite{SethN13}. Rule based systems, which were seminal table extraction techniques, may also be relevant \cite{ShigarovAMPC18}.

\noindent
{\bf Aim and Contribution}

The general purpose is to automatically process invoices to get important data that are contained in these documents and particularly information contained in tables. This work is motivated by a real case application in a full document management system\footnote{This work is conducted in association with the KaliConseil company that is developing its own specific invoices processing system.}. 

We focus on different types of information, such as localisation, tables, dates and actors who are organizations or people identified in the invoice. These fields have been completed after analysis of several invoice models and according to the current requests of the companies questioned. From a general point of view, some important information (not limited to) contained in a invoice can be sketched as follows :
\begin{itemize}
\item \textbf{actors} individual, company or companies involved in the invoice ( customer or supplier).
\item \textbf{addresses} all the addresses contained in the document and, if available,  their types, billing address, delivery or sender, for instance
\item \textbf{dates} the set of dates, specific to the invoice process such as edition date, payment date ...
\item \textbf{tables} tables often presenting the invoiced items, quantities, prices...
\end{itemize}

In this paper, we propose an integrated approach to extract information that are formatted into tables in invoices. The purpose is to extract the whole data, and not only specific labels such as "total price" or "product description", in order to be able to automatically process these data, for instance into a dedicated database that could be used for invoice information retrieval. In our collected data set, the tables are not necessarily drawn using lines to precisely delimit information and they may include missing information. 

Given an invoice, we consider that a table can be detected at two levels : either by detecting visually some characteristic shapes (vertical and/or horizontal lines) or by detecting some structural organization of the tokens of information, aggregated according to rows and columns that may intersect. We refer to this second level as a semantic level since it uses an intrinsic model that tells us what a table should be, even if its graphical frame is not necessary fully present. Our contribution is twofold. On the one hand we propose a complete formalization of the document based on ordering relations that help us to more precisely process a graph-based model of the structure of the document. On the other hand, we combine visual analysis of the document with this more semantic structure to design an efficient table extraction tool.  
 
\noindent
{\bf Organization}

Section \ref{sec:data} introduces the domain of data recognition and our modeling of invoices. Section \ref{sec:relations} presents the ordering relations used in our graph-based approach. Section \ref{sec:patterns} introduces the notion of pattern. Section \ref{sec:algo} presents our new approach to extract tables and Section \ref{sec:results} shows its efficiency with experiments.

\section{Data recognition}
\label{sec:data}

\subsection{Optical Character recognition}

As mentioned in the introduction, the first task is to recognize the text in a scanned document by means of an optical character recognition (OCR) system. Here, we use  Flexicapture \cite{ vlada2010abbyy}\footnote{The choice of Flexicapture is motivated by our practical industrial tool.}. The full description of OCR techniques is not in the scope of this paper, nevertheless we point out some biases of this kind of tool.  

\begin{itemize}
\item Punctuation recognition can be complex for some document. The tool can be configured to change automatically the format of the dates, for example 01/01/1992 will be transformed into 01.01.1992

\item Certain digits are complex to be recognized and can be easily confused with letters. This problem may depend on the font used in the document. For instance, "0oOo" can be confusing for an OCR
\item The quality of the digitization is an important parameter. There is indeed a strong correlation between the quality of the scanned document and the errors made by the OCR system. 
\end{itemize}

This step allows us to generate a searchable PDF that includes images obtained during the digitization and a non visible text layer. This text layer includes the characters and their geographical position in the document. Note the searchable pdf documents are documents whose text element can be selected. 

\subsection{Tokenization}

A token corresponds to a useful semantic unit according to a given target language. The tokenization process is often considered as the pre-processing of any natural language processing system (NLP) \cite{webster1992tokenization}. In our work, we use word level tokenization. Once the OCR processing has been performed, an API, called PDFBox, is used  to extract the text layer from the PDF document, we use this API to tokenize the extracted text, by separating each word by a maximal distance of two character size in the lowest font.


Once the text has been extracted and the positions of the characters have been retrieved, we generate word level tokens by grouping the characters. The tokens include the characters and the coordinates of a box embedding all the characters.
\begin{figure}[ht]
\centering
\includegraphics[width=0.80\textwidth]{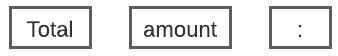}
\caption{Example of obtained tokens}
\label{fig:my_label}
\end{figure}

\subsection{Entity Recognition}


First presented in \cite{marsh1998muc} named entity recognition (NER) consists in the labeling of a text where each string is associated to a person, a location, an organization, a temporality, an amount or a percentage...
Later, NER has evolved to consider more or less labels \cite{sang2003introduction,doddington2004automatic,alfonseca2002unsupervised,evans2003framework}. In our context, we used regEx to perform simple rule-based NER, which is sufficient. Our NER is restricted to a few labels due to our restricted context but when faced to more open domains, more labels are necessary \cite{alfonseca2002unsupervised}. In our case, some tags like header recognition can easily be improved by using a keyword library.


\section{Ordering relations for document representation}
\label{sec:relations}
Using graph-based representation of document has already been explored \cite{ShafaitS10} as mentioned in Introduction. Nevertheless, since we want to address more complex possible structures in the document, we define several relations that will correspond to different abstraction levels in the documents. The purpose of this section is to better define the different abstraction levels that allow us to progressively handle basic tokens of the document to embed them in more complex structures. Hence, we introduce a model based on ordering and partial ordering relation to more precisely define the relationships between information pieces contained in our documents. From now on, we consider that a table is composed of lines and columns. Note we will use the term "line" or "row" indifferently in the following.

\subsection{Basic alignment relations}
Let us consider two tokens $t_1,t_2 \in \textit{T}$, where $\textit{T}$ is the set of tokens obtained after applying the OCR and {\em tokenization} processes described above. Each token $t_i$ is included in a box whose coordinates are respectively $X_{t_i}$ and $Y_{t_i}$. We define two basic binary relations on $\textit{T}$. 
\begin{equation}
r_{horizontal}(t_1,t_2) \Leftrightarrow (|Y_{t_1} - Y_{t_2}| < \tau_Y)
\end{equation}
\begin{equation}
 	r_{vertical}(t_1,t_2) \Leftrightarrow (|X_{t_1} - X_{t_2}| < \tau_X)
\end{equation}
where $\tau_X$ and $\tau_Y$ are two thresholds that are defined to accept some possible tolerance in the vertical and horizontal alignments of the tokens. 

\begin{figure}[hbt!]
\centering
\includegraphics[width=\textwidth]{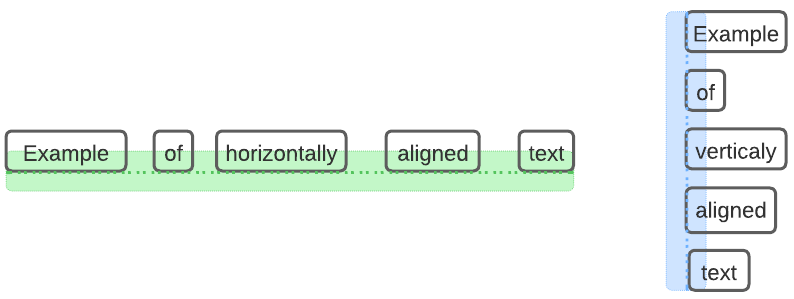}
\caption{Example of recognized relations}
\label{fig:my_label}
\end{figure}

Nevertheless, this approach has some limitations in presence of  texts  that are justified using center or right alignments. Hence, we propose to use interval, inspired by \cite{trapRange}. 
We consider that there is an alignment relation as soon as the interval of positions containing the first and last characters of the tokens have an intersection. For a token $t_i \in \textit{T}$ this interval is denoted $[X_{t_i},endX_{t_i}]$, since $X_{t_i}$ was the position on X axis of the first character. Hence, we get :
 
\begin{equation}
r_{vertical}(t_1,t_2) \Leftrightarrow X_{t_1} \in [X_{t_2},endX_{t_2}] \vee X_{t_2} \in [X_{t_1},endX_{t_1}]
\end{equation}

Let us note that due to the use of intervals, $r_{horizontal}$ and $r_{vertical}$ are not necessarily equivalence relations. Hence, we may have $r_{horizontal}(t_1,t_2)$ and $r_{horizontal}(t_2,t_3)$ and, due for instance to a progressive offset to the bottom of the document, $\not r_{horizontal}(t_1,t_3)$. Hence, alignment relations are computed by means of a dedicated algorithm that uses the coordinates of the tokens that have been extracted by the tasks previously described. This algorithm handles previous above mentioned cases to transform the relations into equivalence relations using thresholds. For instance it will decide that $t_1$ and $t_2$ will be on the same line while $t_3$ will belong to next line.   

Hence, these relations allow us to get an abstract of the relative positions of the token, which is the relevant information for extracting tables. Since these relations are now equivalence relations they induce equivalences classes between tokens. Column and line labels can directly be set from these equivalence classes, one label per class, since we are only interested in determining if tokens are on the same line/column. Using the coordinates of the tokens, we assume that we get a totally  ordered set of line labels $(\textit{L},<)$ and  a totally  ordered set of columns labels $(\textit{C},<)$.   

\subsection{Ordering relations on tokens}
We consider now that a document is a set of labeled token $\textit{T}$ such that each token $t \in  \textit{T}$ has a line label $l(t) \in \textit{L}$ and a column label $c(t) \in \textit{C}$. We use the total ordering relations previously defined on labels to get partial ordering relations on tokens.

Given the set of labeled tokens $\textit{T}$ we define two partially ordered sets $(\textit{T},\preceq_{l})$ and $(\textit{T},\preceq_{c})$ where 
\begin{itemize}
\item $t \preceq_{l} t'$ iff $l(t)<l(t')$ and $c(t)=c(t')$ 

 \item $t \preceq_{c} t'$ iff $c(t)<c(t')$ and $l(t)=l(t')$ 
\end{itemize}

We may now precisely define the concept of lines and columns. 

\textbf{Lines /} A line is any subset $L \subseteq \textit{T}$, such that $(L,\preceq_{l})$ is a totally ordered (linearly ordered). $(\textit{L},\subseteq)$ is the partially ordered set of all possible lines.

A full line is a subset $L \subseteq \textit{L}$ such that $L$ is a maximal element of $(\textit{L},\subseteq)$. Note that if $L$ is a full line then $sup_{l}(L) \in L$ and $inf_{l}(L) \in L$.

\textbf{Column /}
A column is any subset $C \subseteq \textit{T}$, such that $(C,\preceq_{c})$ is a totally ordered set. $(\textit{C},\subseteq)$ is the partially ordered set of all possible columns.

A full column is a subset $C \subseteq \textit{C}$ such that $C$ is a maximal element of $(\textit{C},\subseteq)$. Note that if $C$ is a full column then $sup_{c}(C) \in L$ and $inf_{c}(C) \in L$.

\subsection{Tables through different levels of abstraction}
\label{sec:tables}
Hence, a table is a set of tokens composed by lines and columns that share elements. Let us define a table ordering on $\textit{T}$ as $\preceq_{t} = (\preceq_{l} \cup \preceq_{c})^*$, i.e. the transitive closure of the union of the two previous ordering relations on lines and columns. Let us note that for two tokens $t,t' \in \textit{T}$, $t \preceq_{t} t'$ iff $l(t)\leq l(t')$ and $c(t) \leq c(t')$.

\textbf{Table /} A table is any subset $T \subseteq \textit{T}$, such that $(T,\preceq_{t})$ is a complete lattice. As a consequence table $T \subseteq \textit{T}$ satisfies $sup_{t}(T) \in T$ (its greatest element $\top_T$) and $inf_{t}(T) \in T$ (its least element $\bot_T$). A table can be then defined by these two elements $T=(\bot_T,\top_T)$. 

As a consequence, in a table $T$, given any two tokens $t_1,t_2 \in T$, $sup_{t}(\{t_1,t_2\})$ and $inf_{t}(\{t_1,t_2\})$ exist and constitute the smallest sub-table $T'=(sup_{t}(\{t_1,t_2\},inf_{t}(\{t_1,t_2\})$ that contains $t_1$ and $t_2$. 

Let us consider $(\textit{T},\subseteq)$ a the partially ordered set of all possible tables. A full table is a subset $T \subseteq \textit{T}$ such that $T$ is a maximal element of $(\textit{T},\subseteq)$.

\textbf{well-formed full table /} A full table $T$ is well-formed iff : 
\begin{itemize}
\item $T$ can be partitioned into a set $\{ L_1,\cdots,L_n\}$ of full lines, such that $sup_t(T)=sup_l(L_n)$ and $inf_t(T)=inf_l(L_1)$ 
\item $T$ can be partitioned into a set $\{ C_1,\cdots,C_m\}$ of full columns, such that $sup_t(T)=sup_c(C_n)$ and $inf_t(T)=inf_c(C_1)$
\item
$\forall 1\leq i \leq n,1\leq j \leq m, |L_i \cap C_j|=1$
\end{itemize}

Since we want to detect general tables, with possibly missing cells, we may turn any table into a well-formed full table by adding empty cells, i.e. new tokens. 

\begin{itemize}
\item Lines:
\begin{enumerate}
   \item  $T$ cannot be partitioned into a set $\{ L_1,\cdots,L_n\}$ since two lines $L_i$ and $L_j$ intersect: each element of $L_i \cap L_j$ is duplicated such that $L_i \cap L_j = \emptyset$: add then the corresponding ordering relations.
   \item 
   A line $L_i$ is not a full line. There exists $t \in \textit{T}$ such that $L_i \cup \{t\}$ is a full line. Add to each line $L_j, j\neq i$ a new element $t_j$ such that $\{t\} \cup \bigcup_j \{t_j\}$ is a full column. Add then the corresponding ordering relations.
\end{enumerate}
\item Columns: same processes
\end{itemize}

\begin{figure}[ht]
\centering
\includegraphics[width=\textwidth]{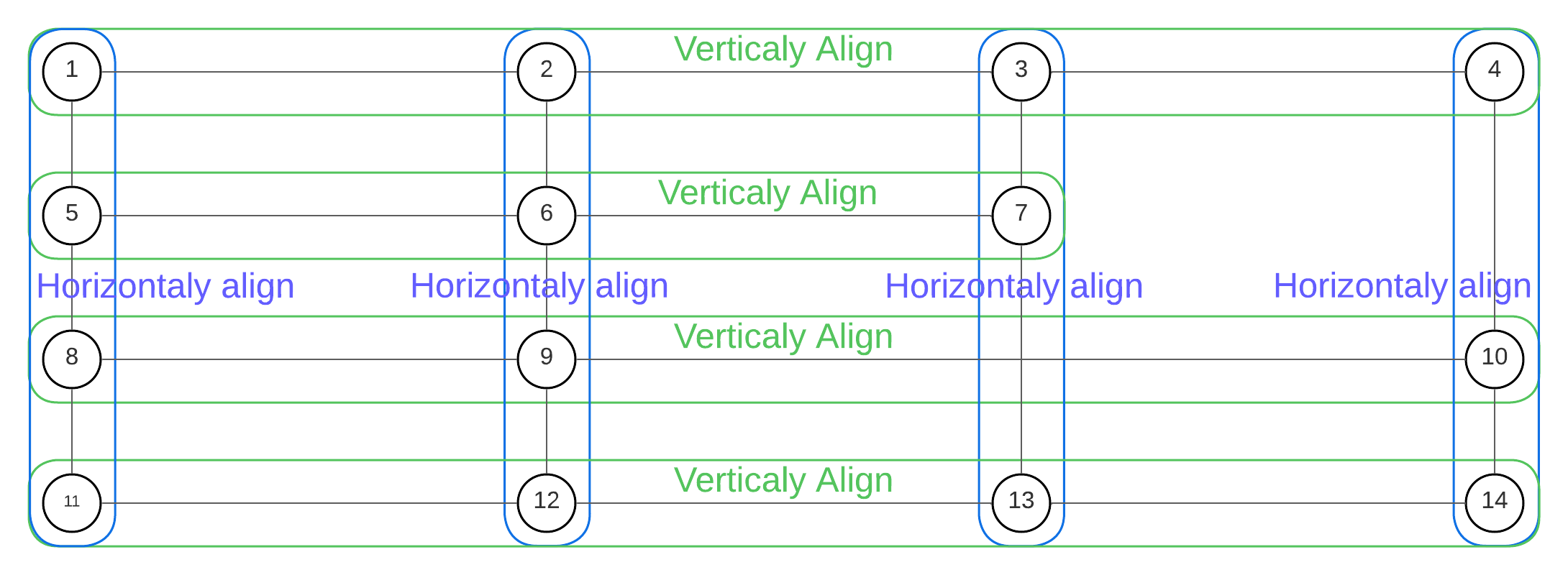}
\caption{Table with empty cells}
\label{fig:empty_cells}
\end{figure}

Figure \ref{fig:empty_cells} shows an example of a table with missing cells. 

Note that this repairing process will be performed of initial set of tokens to get better candidate lines and columns for obtaining better tables in the documents. 

At this stage, we have a clear modelling of lines, columns and possibly resulting tables. In order to reach a higher level of abstraction and to manipulate directly lines and columns, we extend our ordering relations on tokens to lines and columns. Let us consider $\textit{L}^+\subseteq \textit{L}$ (resp. $\textit{C}^+\subseteq \textit{C}$) the set of full lines (resp. columns). According to the previous definitions, note that $\textit{L}^+$ and $\textit{C}^+$ are partitions of $\textit{T}$. We consider now the two orderings $(\textit{C}^+,\preceq_{c})$ and $(\textit{L}^+,\preceq_{l})$ that are the canonical extensions of $(\textit{T},\preceq_{c})$ and $(\textit{T},\preceq_{l})$. Note that here, due to the fact that we consider separately full lines and full columns, these orders are total orders. Searching for a table corresponds in searching for a subset of lines and columns. According to previous definitions, a table $T$ can be defined by a couple of sets $T=(L,C)$, such that $L \subseteq \textit{L}^+$ and $C \subseteq \textit{C}^+$. The structure of the table is then defined by the possible insertions between lines and columns (i.e, common tokens).

\subsection{From orders to graphs}
\label{sec:order_to_graph}

Our formalism allows us to get a clear an general abstract view of a table in a document, where information has been grouped into basic tokens. Now since we want to get operational tools for computing tables, we turn our  orders into a graph representation using the classic Hasse diagram (i.e. transitivity is not represented to simplify the graph). 

We consider a first directed graph $(\textit{L}^+,E_L$ where $(l_i,l_j) \in E_L$ iff $l_i <_l l_j$ and $\nexists l_k \textit{L}^+, l_i <_l l_k <_l l_j$. We consider a similar graph $(\textit{C}^+,E_C$ for the columns with similar properties). We add an undirected graph $(\textit{L}^+ \cup \textit{C}^+, E_{\cap}$ where $[l_i,c_j] \in  E_{\cap}$ iff $l_i \cap c_j \neq \emptyset$). 

The final graph is thus a graph that gather the three previous graphs whose edges have different types according to their semantics (lines and columns ordering or intersections) $(\textit{L}^+ \cup \textit{C}^+,E_L\cup E_C \cup E_{\cap})$. Using this graph representation, a table corresponds to a particular sub-graph in the graph that represents the whole document. This graph is process into a solver that is dedicated to sub-graph search \cite{mccreesh2018subgraph}, allowing our different types of edges. Note that a table may contain empty cells. Hence, we introduce possible empty nodes in our graph representation. Now we have to introduce patterns that  correspond to general possible structures of tables that we search for in our graph. 

\section{Search Using Patterns}
\label{sec:patterns}

The tables contained in a document can have many different forms. Each of them can be defined with mandatory characteristics (existence of a complete first row, existence of a cell in a specified corner ...). We use patterns represented by graphs, using the previously introduced formalism,  to represent these characteristics. A pattern represent indeed a set of similar tables. Searching for valid tables for a given pattern in a graph therefore leads to the graph isomorphism problem. This well known NP-intermediate problem is solved used the solving process detailed in Section \ref{sec:order_to_graph}.

\subsection{Tables generalized as pattern}

A pattern represents a set of tables sharing the same characteristics. As characteristics, the number of rows and columns are necessary for pattern definition. Remind that according to Section \ref{sec:order_to_graph}, the graph corresponding to a pattern is built: 
\begin{itemize}
\item Each line and column is represented by a vertex in the pattern. 
\item Vertical and horizontal alignment relations are represented by directed arcs. 
\item The presence of a cell at the intersection of a row and a column is represented by an non-directed edge between the row vertex and the column vertex.
\end{itemize}

\begin{figure*}[hbt!]
\begin{center}
\begin{tabular}{c c | c c}

\includegraphics[width=0.20\textwidth]{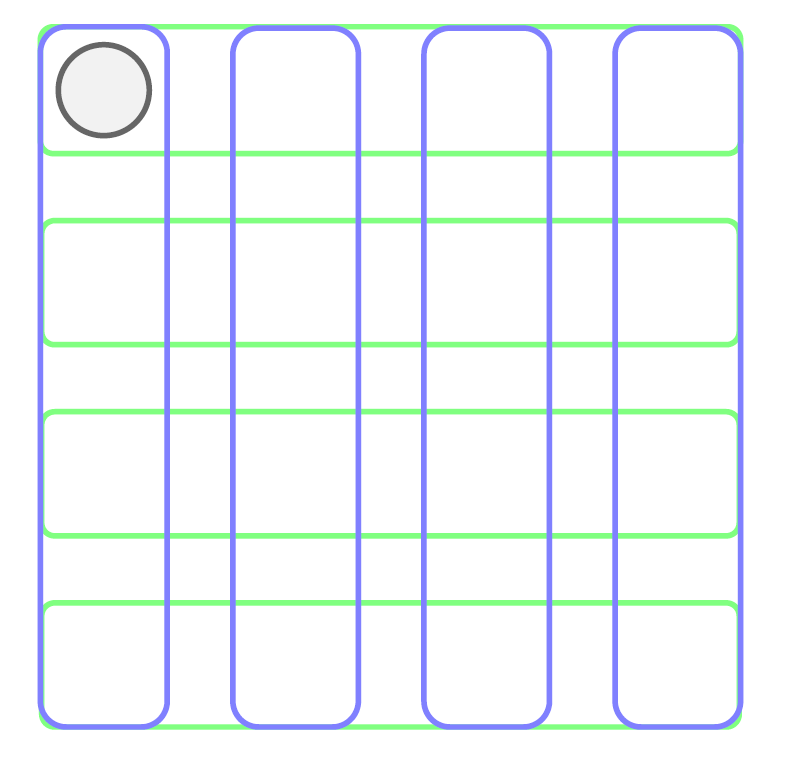} &
\includegraphics[width=0.25\textwidth]{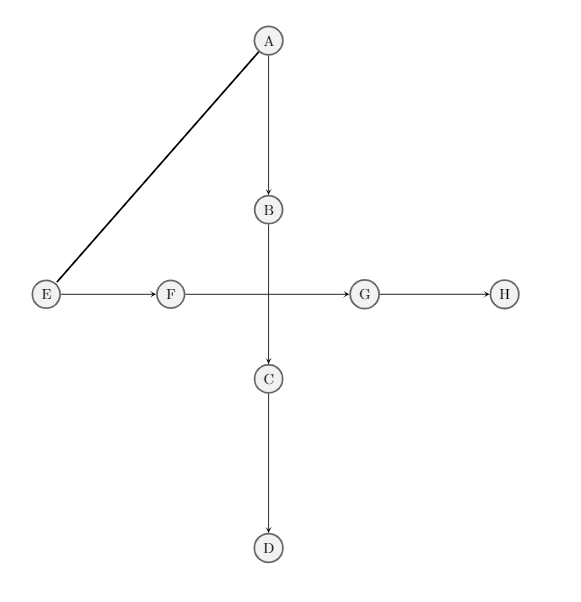}&
\includegraphics[width=0.20\textwidth]{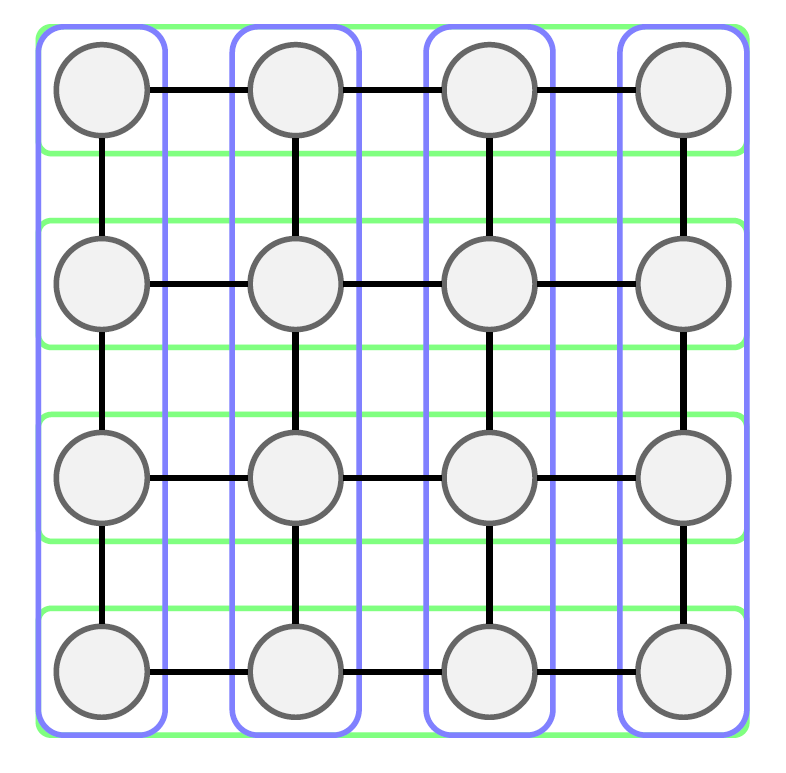} &
\includegraphics[width=0.25\textwidth]{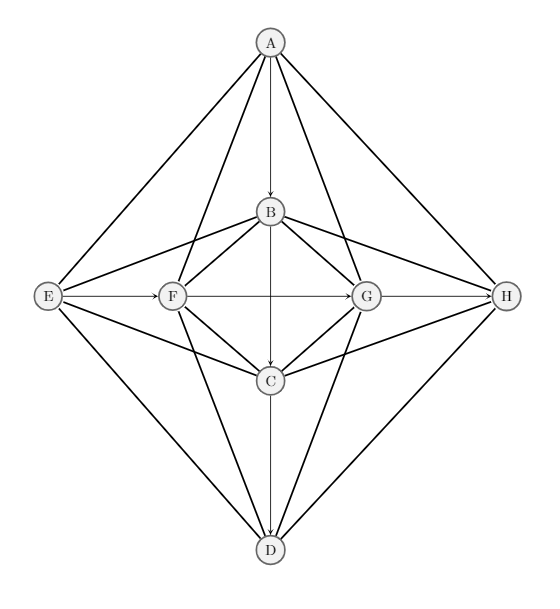}\\
\multicolumn{2}{c|}{Corner Left Top}&
\multicolumn{2}{c}{Full grid}\\
\hline

\includegraphics[width=0.20\textwidth]{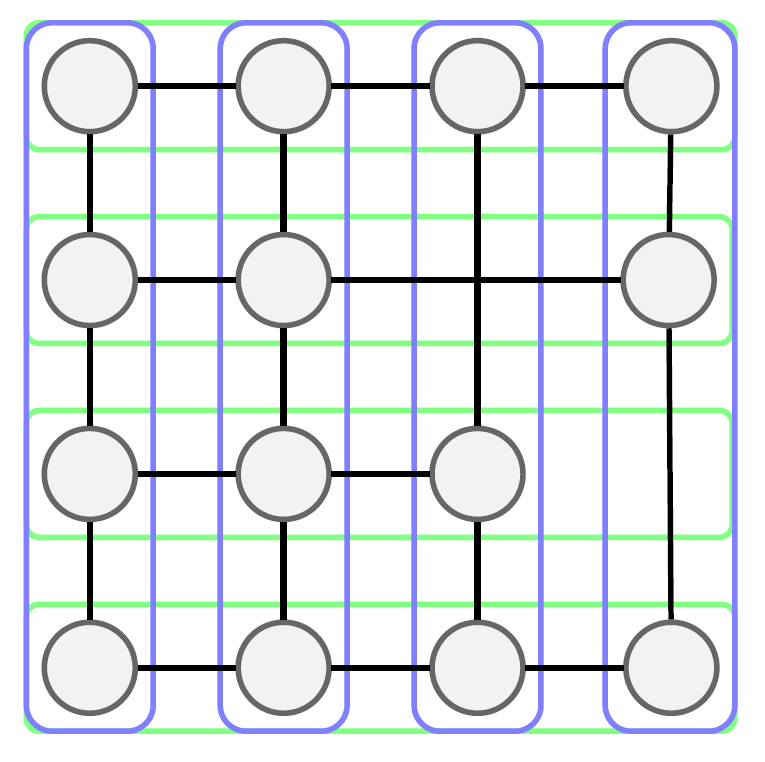} &
\includegraphics[width=0.25\textwidth]{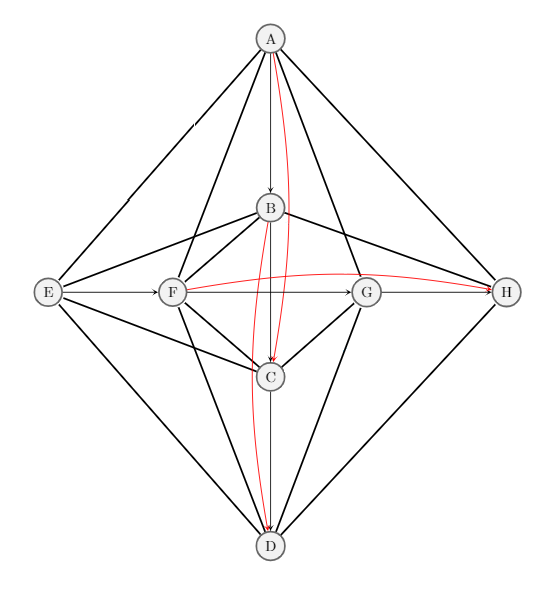}& 
\includegraphics[width=0.20\textwidth]{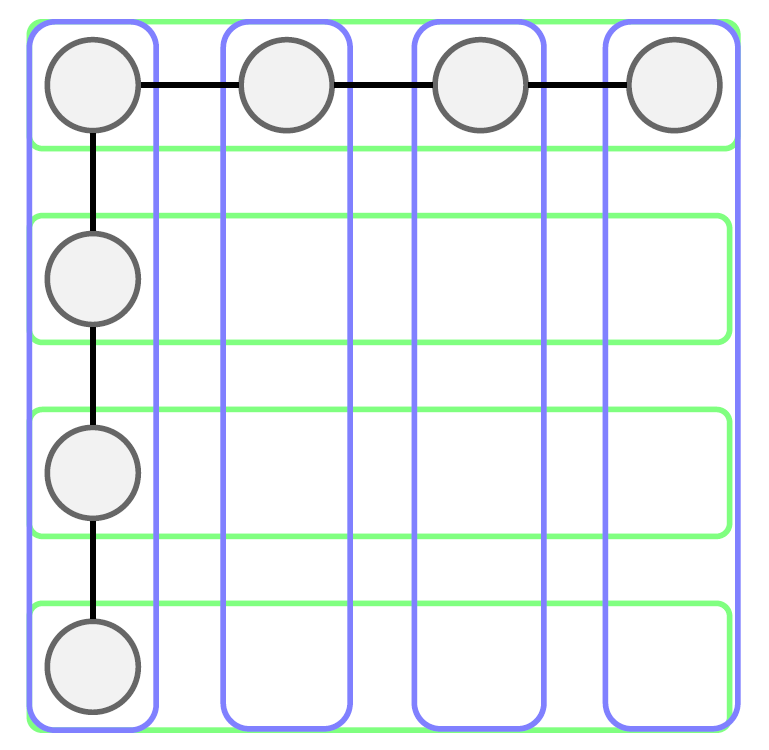} &
\includegraphics[width=0.25\textwidth]{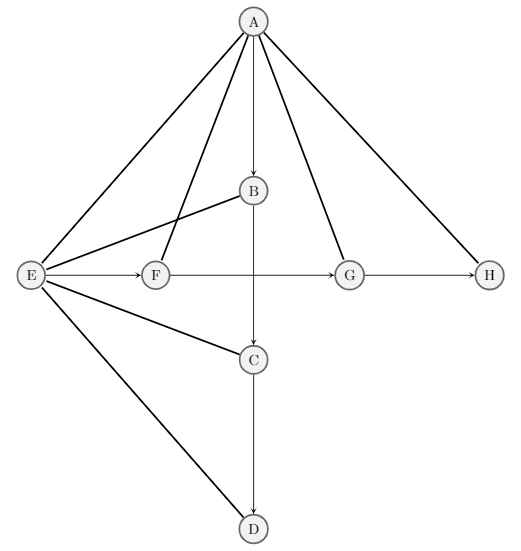} \\
\multicolumn{2}{c|}{Missing cells}&
\multicolumn{2}{c}{Border Left Top}\\
\end{tabular}
\end{center}
\caption{Different patterns for a $4 \times 4$ table.}
\label{fig:patterns}
\end{figure*}

Figure \ref{fig:patterns} presents patterns with 4 lines and 4 columns respectively represented by vertices A, B, C, D and E, F, G, H. Each pattern is presented with its table diagram on the right and its corresponding graph on the left. These four patterns are issued from practical needs and correspond to common patterns that are encountered in invoices. They have some specific properties detailed below : 
\begin{itemize}
\item \textbf{Corner Left Top /} We can see that an arc in the pattern induces the presence of a token at the intersection the first line (from top to bottom) and the first column (from left to right). No other common cell is required to be shared by the remaining of the table. 
   \item \textbf{Full grid /} This table corresponds to a well formed full table as formally defined in Section \ref{sec:tables}.
\item \textbf{Missing cells /} As explained in Section \ref{sec:tables}, some tables may contain empty cells. This pattern shows that in this case some alignment relations may be missing. Nevertheless such a table is still valid since it can be completed into a full table by introducing empty cells. In our graph model we are able to generate a suitable pattern for searching for such a table. 
\item \textbf{Border Left Top /} This pattern is based on the assumption that there is at less one complete line to the top and one complete column to the right in our table.
\end{itemize}

Let us note that an undirected edge between a line and a column represents a constraint of presence of a token at their intersection but the absence or edge does not necessarily induce the non-presence of an intersection. In  the corresponding table schemes of Figure \ref{fig:patterns}, this mandatory presence of a token is indicated by a light grey circle. In these scheme, lines are represented by blue rectangles that may have possibly empty cells, which correspond to the possible absence of token at this place. Columns are represented by green rectangles, following the same principles. 


For instance, given a well formed table with all cell filled, our "full grid" pattern is able to recognize the table. But in case of a table containing empty cells, the "full grid" pattern does not match because it is too restrictive.

Let us note that the patterns presented is Figure \ref{fig:patterns} can be defined for any arbitrary number of lines and columns, resulting in vertices in our graph. These two parameters define the size of the pattern. 

\subsection{Pattern-based Algorithm for Tables Detection}

As already explained, searching for a pattern in the whole document modelled as a graph  corresponds to solve the sub-graph isomorphism problem \cite{Ullmann76, Prosser93}. 

Given a pattern type, our objective is to find the pattern of largest size, i.e, maximal number of lines/columns. To achieve this incremental pattern search, we use the algorithm proposed in \cite{mccreesh2018subgraph} to find a pattern with a given size and, as long as no pattern is found, we continue the search by progressively decreasing the size of the pattern.

\section{the visual approach and the graph-based approach} 
\label{sec:algo}
Due to the complexity of the sub-graph search problem, the graph-based approach requires computational resources to identify tables. Nevertheless, in some cases, the detection of horizontal and vertical lines drawn in document may be sufficient to identify tables structures, when they are obvious. This detection can be performed by a  visual processing tools in an inexpensive way. Combining the performance of our graph based approach and the speed of the visual detection help us to make our approach more robust  for the different documents that can be processed.

\subsection{Visual Processing}

The approach to exploit the token placement in documents does not use detection of visual elements such as lines (in the graphic sense). Hence, we propose to integrate specific tools for visual processing in order to enrich the information of the processed document. The OpenCV library \cite{bradski2000intel} is currently one of the most used when dealing with information extraction from documents. Several approaches are available in this library but those dealing with lines and contours are the most adapted to our problem. The sequence of several treatments (RGB2Gray, threshold, boudingRect ...) permits the detection of graphical information.

Other approaches such as the Canny algorithm \cite{canny1986computational} allow to return visual information but it is inappropriate for tables. Indeed, the precision and the number of returned information make this approach too expensive for our problem.

\subsection{Strategies to explore the graph-based representation}
\label{sec:strategy}

OpenCV provides us with local information on the document. We have developed a tool that takes into account this information from OpenCV and aggregates it to extract areas that  seem to visually correspond to a table. In fact, we use the output of OpenCV as a surrogate approximation of the shape of the document to evaluate if a table is likely to be expected or not. We define several possible strategies for this approximation. 

\begin{itemize}
\item \textbf{Empty} No information is returned by the visual processing. The document is therefore fully processed with the graph-based approach to check the presence of a table.  
\item \textbf{Area} An area is identified by the visual processing but no additional information is provided. Only this area is processed with the graph-based approach. 
\item \textbf{Column} An area is detected by visual processing and vertical lines allow the identification of columns. Only this area is processed with the graph-based approach but the identified columns are fixed in advance. 
\item \textbf{Grid} An area is detected by visual processing with a full grid allowing to identify a table. Only this area is processed with the graph-based approach but the identified columns and lines are fixed in advance.
\end{itemize}

The strategy selection is therefore completely dependent on the visual processing result. The quality of the table detection remains the same whatever the strategy but the computational power used is minimized because the cost of the visual processing is negligible compared to that of the graph-based approach applied to the full document.

\subsection{Overall process}

\begin{figure}[ht]
\centering
   \includegraphics[width=0.5\textwidth]{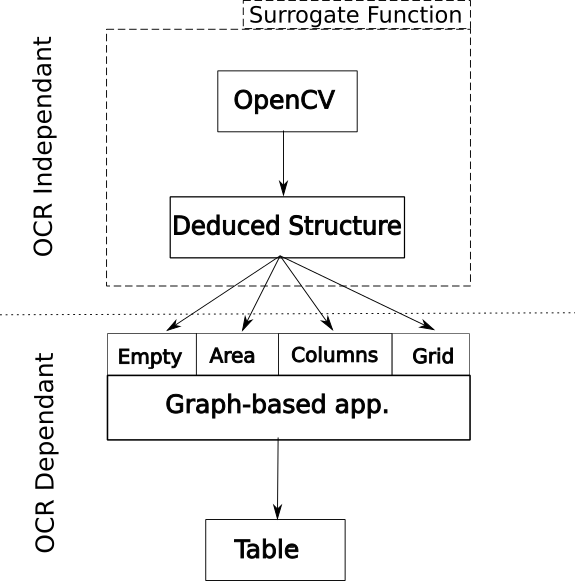}
\caption{Overall process of our approach.}
\label{fig:opencv}
\end{figure}

The overall process we used for our experiments includes the use of OpenCV, our OpenCV based estimation tool and our graph-based approach. Figure \ref{fig:opencv} describes this process graphically.
A first step is OCR-independent since OpenCV and our estimation tool only use graphical information. As already mentioned, this step can be seen as a surrogate function because it allows us to characterise very efficiently the area to be treated and its associated features.
The second step introduces more semantic information, using tokens, and it is very dependent on the quality of the OCR. The strategy of the graph-based approach used depends thus on the first step, involving more or less computational resources. 

\section{Experimental Evaluation}
\label{sec:results}
 
To test our approach we have to be able to evaluate and compare the tables that have been recognized. Hence, we propose a fitness function and a set of benchmark instances. Using the fitness function on our benchmark will allows us to assess the efficiency of our approach. 
 
\subsection{Evaluation of an output table}
 
It is really hard to evaluate automatically the tables provide by a solver. 
The evaluation of the output of our method is a complex problem and it is indeed difficult to evaluate table extraction mechanisms. We need to have a set of labelled invoices. Hence, we have manually labelled the graphs corresponding to our invoices and use the following function to compute the distance between the optimal tables (graph manually labelled) and the output tables. Given an output table $t$, its table accuracy $t_a$ (fitness) is evaluated as :

\begin{equation}
t_a = \left( 1 - \left|\frac{t_w-t_f}{t_w+t_f}\right|\right)\left( 1 - \left|\frac{t_w-t_t}{t_w+t_t}\right|\right)
\end{equation}

where $t_w$ is the number of expected token from labeled result, $t_f$ is the number of token in output result and $t_t$ is the number of token correctly found. Note that the left part of the formula corresponds to the precision and the right part to the recall. This formula is similar to the work presented by Bissan A. et al \cite{audeh2015mor} and the notion of totality. Here, we consider $\frac{t_w-t_t}{t_w+t_t}$ as our normalized recall.

To get a better evaluation of the structure, we use a specific evaluation at the column level. A column with the expected number of tokens in each cells is called complete and incomplete otherwise. $c_c$ is number of complete columns found and $c_p$ of partial columns found. $c_w$ is the number of expected columns while $c_t$ is the total number of columns found in the output table.  

\begin{equation}
c_a = \frac{2c_c+c_p}{3c_w} \left( 1 - \left|\frac{c_w-c_t}{c_w+c_t}\right|\right)
\end{equation}

The coefficients have been set by experiments. We use the same evaluation for the lines of the tables to get $l_a$. 

The previous evaluations are aggregated into a single fitness value for the output table as 

\label{fitness}
$$f = \frac{t_a + c_a * l_a}{2}$$

This function is not intended to be used in the overall process. We just use it to assess the efficiency of the different proposed approaches.
 
\subsection{Benchmark instances}

The set of invoices comes from a real set of data issued from a practical application in a company. These recent invoices are issued from data sets provided by customers that need to process them. Hence, we cannot make them available but our solution is available for all type of invoice. To cover a sufficiently large panel of invoices, hundreds of documents have been examined. We have selected 21 instances for which we have tested the 4 different approaches. Each instance is classified according to a human expertise. Hence our benchmark set covers very different types of tables that are contained in the invoices. 

We do not use typical benchmarks for table recognition, because our main goal is not only to locate a table, but to extract the data which may require a reliable level of OCR processing and therefore a high document quality. Classical benchmarks are often poorly digitized and present scanning noise, which is indeed a challenge in this case. Nevertheless, our work does not deal with the performance of the OCR or image processing techniques and such data sets do not seem relevant to us. Moreover, our chosen instances are more representative for the real application case we want to address. 

\subsection{Experimental Results}

Our experimental protocol is the following. We consider different table extraction processes based on the different stages and strategies we have previously defined. Since the choice of the pattern is an important parameter for our method, we have tested different patterns and, as expected, there is no overall better pattern for all the instances. Hence, we only use  a pattern similar to  the Border Left Top pattern presented in Figure \ref{fig:patterns}. Running time is limited to 300 seconds.

Our aim is to show that the process that includes the surrogate evaluation of the document and a suitable choice of the graph-based strategy provides efficient results. In Table \ref{tab:results} we first report the results of our augmented OpenCV based function that estimate the type of table that is expected in the document. Then, we report each graph exploration strategy according to Section \ref{sec:strategy}. For each possible choice, results in term of score (fitness function presented Section \ref{fitness}) and running time for all the graph-based approaches are given. For the "Graph-based approach" part, the first four column use a single strategy despite the result of the OpenCV based analysis. The dynamic strategy corresponds thus to the choice of the suitable graph search strategy according to the surrogate estimation. Note that all running times are in seconds and even the maximum running time is reached, the best table so far found is returned and then a score can be computed.

\begin{table*}[hbt!]
\resizebox{0.95\textwidth}{!}{
\begin{tabular}{|cl|ll|cccccccccc|}
\hline
\multicolumn{2}{|c|}{\multirow{2}{*}{Instances}}                                           & \multicolumn{2}{c|}{\multirow{2}{*}{\begin{tabular}[c]{@{}c@{}}OpenCV + \\ Deduced Structures\end{tabular}}} & \multicolumn{10}{c|}{Graph-based approach}                                                                                                                                                                                                                                                                                                                                                         \\ \cline{5-14} 
\multicolumn{2}{|c|}{}                                                                     & \multicolumn{2}{c|}{}                                                                                        & \multicolumn{2}{c|}{Grid}                                                       & \multicolumn{2}{c|}{Column}                                                     & \multicolumn{2}{c|}{Area}                                                       & \multicolumn{2}{c|}{Empty}                                                      & \multicolumn{2}{c|}{Dynamic}                               \\ \hline
\multicolumn{1}{|c|}{\multirow{2}{*}{Family}} & \multicolumn{1}{c|}{\multirow{2}{*}{Name}} & \multicolumn{1}{c|}{Score}                       & \multicolumn{1}{c|}{\multirow{2}{*}{Time}}                & \multicolumn{1}{c|}{Score}         & \multicolumn{1}{c|}{\multirow{2}{*}{Time}} & \multicolumn{1}{c|}{Score}         & \multicolumn{1}{c|}{\multirow{2}{*}{Time}} & \multicolumn{1}{c|}{Score}         & \multicolumn{1}{c|}{\multirow{2}{*}{Time}} & \multicolumn{1}{c|}{Score}         & \multicolumn{1}{c|}{\multirow{2}{*}{Time}} & \multicolumn{1}{c|}{Score}         & \multirow{2}{*}{Time} \\
\multicolumn{1}{|c|}{}                        & \multicolumn{1}{c|}{}                      & \multicolumn{1}{c|}{(surrogate)}                 & \multicolumn{1}{c|}{}                                     & \multicolumn{1}{c|}{(fitness)}     & \multicolumn{1}{c|}{}                      & \multicolumn{1}{c|}{(fitness )}    & \multicolumn{1}{c|}{}                      & \multicolumn{1}{c|}{(fitness)}     & \multicolumn{1}{c|}{}                      & \multicolumn{1}{c|}{(fitness )}    & \multicolumn{1}{c|}{}                      & \multicolumn{1}{c|}{(fitness)}     &                       \\ \hline
\multicolumn{1}{|c|}{\multirow{6}{*}{Column}} & Mod\_1                                     & \multicolumn{1}{l|}{Column}                      & \multicolumn{1}{r|}{1.37}                                                      & \multicolumn{1}{r|}{0.01}          & \multicolumn{1}{r|}{3.43}                  & \multicolumn{1}{r|}{\textbf{1.00}} & \multicolumn{1}{r|}{\textbf{3.42}}         & \multicolumn{1}{r|}{0.43}          & \multicolumn{1}{r|}{2.69}                  & \multicolumn{1}{r|}{0.35}          & \multicolumn{1}{r|}{17.69}                 & \multicolumn{1}{r|}{\textbf{1.00}} & \multicolumn{1}{r|}{\textbf{3.42}}         \\ \cline{2-14} 
\multicolumn{1}{|c|}{}                        & Mod\_2                                     & \multicolumn{1}{l|}{Column}                      & \multicolumn{1}{r|}{2.19}                                                    & \multicolumn{1}{r|}{0.00}          & \multicolumn{1}{r|}{300.01}                & \multicolumn{1}{r|}{\textbf{0.51}} & \multicolumn{1}{r|}{\textbf{300.02}}       & \multicolumn{1}{r|}{0.41}          & \multicolumn{1}{r|}{300.01}                & \multicolumn{1}{r|}{0.05}          & \multicolumn{1}{r|}{300.02}                & \multicolumn{1}{r|}{\textbf{0.51}} & \multicolumn{1}{r|}{\textbf{300.02}}       \\ \cline{2-14} 
\multicolumn{1}{|c|}{}                        & Mod\_3                                     & \multicolumn{1}{l|}{Column}                      & \multicolumn{1}{r|}{2.61}                                                     & \multicolumn{1}{r|}{0.00}          & \multicolumn{1}{r|}{4.85}                  & \multicolumn{1}{r|}{\textbf{1.00}} & \multicolumn{1}{r|}{\textbf{4.96}}         & \multicolumn{1}{r|}{0.60}          & \multicolumn{1}{r|}{3.10}                  & \multicolumn{1}{r|}{0.06}          & \multicolumn{1}{r|}{10.30}                 & \multicolumn{1}{r|}{\textbf{1.00}} & \multicolumn{1}{r|}{\textbf{4.96}}         \\ \cline{2-14} 
\multicolumn{1}{|c|}{}                        & Mod\_4                                     & \multicolumn{1}{l|}{Grid}                        & \multicolumn{1}{r|}{2.21}                                                      & \multicolumn{1}{r|}{0.11}          & \multicolumn{1}{r|}{6.80}                  & \multicolumn{1}{r|}{\textbf{1.00}} & \multicolumn{1}{r|}{\textbf{7.12}}         & \multicolumn{1}{r|}{0.48}          & \multicolumn{1}{r|}{9.10}                  & \multicolumn{1}{r|}{0.01}          & \multicolumn{1}{r|}{300.03}                & \multicolumn{1}{r|}{1.00}          & \multicolumn{1}{r|}{7.12}                 \\ \cline{2-14} 
\multicolumn{1}{|c|}{}                        & Mod\_5                                     & \multicolumn{1}{l|}{Column}                      & \multicolumn{1}{r|}{2.01}                                                    & \multicolumn{1}{r|}{0.01}          & \multicolumn{1}{r|}{48.27}                 & \multicolumn{1}{r|}{\textbf{1.00}} & \multicolumn{1}{r|}{\textbf{47.17}}        & \multicolumn{1}{r|}{0.52}          & \multicolumn{1}{r|}{45.77}                 & \multicolumn{1}{r|}{0.44}          & \multicolumn{1}{r|}{64.90}                 & \multicolumn{1}{r|}{\textbf{1.00}} & \multicolumn{1}{r|}{\textbf{47.17}}       \\ \cline{2-14} 
\multicolumn{1}{|c|}{}                        & Mod\_6                                     & \multicolumn{1}{l|}{Column}                      & \multicolumn{1}{r|}{2.37}                                                      & \multicolumn{1}{r|}{0.56}          & \multicolumn{1}{r|}{4.72}                  & \multicolumn{1}{r|}{\textbf{1.00}} & \multicolumn{1}{r|}{\textbf{4.96}}         & \multicolumn{1}{r|}{1.00}          & \multicolumn{1}{r|}{6.90}                  & \multicolumn{1}{r|}{0.01}          & \multicolumn{1}{r|}{10.71}                 & \multicolumn{1}{r|}{\textbf{1.00}} & \multicolumn{1}{r|}{\textbf{4.96}}     \\ \hline
\multicolumn{1}{|c|}{\multirow{5}{*}{Empty}}  & Mod\_1                                     & \multicolumn{1}{l|}{Area}                        & \multicolumn{1}{r|}{2.71}                                                 & \multicolumn{1}{r|}{0.00}          & \multicolumn{1}{r|}{11.28}                 & \multicolumn{1}{r|}{0.00}          & \multicolumn{1}{r|}{24.38}                 & \multicolumn{1}{r|}{0.00}          & \multicolumn{1}{r|}{12.63}                 & \multicolumn{1}{r|}{\textbf{0.30}} & \multicolumn{1}{r|}{\textbf{8.05}}         & \multicolumn{1}{r|}{0.00}          & \multicolumn{1}{r|}{8.05}                  \\ \cline{2-14} 
\multicolumn{1}{|c|}{}                        & Mod\_2                                     & \multicolumn{1}{l|}{Empty}                       & \multicolumn{1}{r|}{0.45}                                                   & \multicolumn{1}{r|}{0.16}          & \multicolumn{1}{r|}{2.64}                  & \multicolumn{1}{r|}{0.32}          & \multicolumn{1}{r|}{3.22}                  & \multicolumn{1}{r|}{0.32}          & \multicolumn{1}{r|}{2.95}                  & \multicolumn{1}{r|}{\textbf{0.51}} & \multicolumn{1}{r|}{\textbf{2.31}}         & \multicolumn{1}{r|}{\textbf{0.00}} & \multicolumn{1}{r|}{\textbf{2.31}}        \\ \cline{2-14} 
\multicolumn{1}{|c|}{}                        & Mod\_3                                     & \multicolumn{1}{l|}{Area}                        & \multicolumn{1}{r|}{0.98}                                                      & \multicolumn{1}{r|}{0.00}          & \multicolumn{1}{r|}{2.78}                  & \multicolumn{1}{r|}{0.00}          & \multicolumn{1}{r|}{2.66}                  & \multicolumn{1}{r|}{0.00}          & \multicolumn{1}{r|}{2.52}                  & \multicolumn{1}{r|}{\textbf{0.48}} & \multicolumn{1}{r|}{\textbf{4.44}}         & \multicolumn{1}{r|}{0.00}          & \multicolumn{1}{r|}{4.44}                  \\ \cline{2-14} 
\multicolumn{1}{|c|}{}                        & Mod\_4                                     & \multicolumn{1}{l|}{Area}                        & \multicolumn{1}{r|}{0.98}                                                     & \multicolumn{1}{r|}{0.00}          & \multicolumn{1}{r|}{2.55}                  & \multicolumn{1}{r|}{0.00}          & \multicolumn{1}{r|}{2.77}                  & \multicolumn{1}{r|}{0.00}          & \multicolumn{1}{r|}{2.62}                  & \multicolumn{1}{r|}{\textbf{0.50}} & \multicolumn{1}{r|}{\textbf{7.72}}         & \multicolumn{1}{r|}{0.00}          & \multicolumn{1}{r|}{7.72}                  \\ \cline{2-14} 
\multicolumn{1}{|c|}{}                        & Mod\_5                                     & \multicolumn{1}{l|}{Area}                        & \multicolumn{1}{r|}{1.67}                                                     & \multicolumn{1}{r|}{0.03}          & \multicolumn{1}{r|}{5.21}                  & \multicolumn{1}{r|}{0.00}          & \multicolumn{1}{r|}{5.35}                  & \multicolumn{1}{r|}{0.00}          & \multicolumn{1}{r|}{4.77}                  & \multicolumn{1}{r|}{\textbf{0.35}} & \multicolumn{1}{r|}{\textbf{9.56}}         & \multicolumn{1}{r|}{0.00}          & \multicolumn{1}{r|}{9.56}                  \\ \hline
\multicolumn{1}{|c|}{\multirow{5}{*}{Grid}}   & Mod\_1                                     & \multicolumn{1}{l|}{Grid}                        & \multicolumn{1}{r|}{0.88}                                                    & \multicolumn{1}{r|}{1.00}          & \multicolumn{1}{r|}{2.75}                  & \multicolumn{1}{r|}{1.00}          & \multicolumn{1}{r|}{2.68}                  & \multicolumn{1}{r|}{\textbf{1.00}} & \multicolumn{1}{r|}{\textbf{2.62}}         & \multicolumn{1}{r|}{0.43}          & \multicolumn{1}{r|}{2.69}                  & \multicolumn{1}{r|}{1.00}          & \multicolumn{1}{r|}{2.75}                \\ \cline{2-14} 
\multicolumn{1}{|c|}{}                        & Mod\_2                                     & \multicolumn{1}{l|}{Grid}                        & \multicolumn{1}{r|}{0.56}                                                   & \multicolumn{1}{r|}{1.00}          & \multicolumn{1}{r|}{2.92}                  & \multicolumn{1}{r|}{1.00}          & \multicolumn{1}{r|}{2.96}                  & \multicolumn{1}{r|}{\textbf{1.00}} & \multicolumn{1}{r|}{\textbf{2.88}}         & \multicolumn{1}{r|}{0.27}          & \multicolumn{1}{r|}{6.35}                  & \multicolumn{1}{r|}{1.00}          & \multicolumn{1}{r|}{2.92}               \\ \cline{2-14} 
\multicolumn{1}{|c|}{}                        & Mod\_3                                     & \multicolumn{1}{l|}{Grid}                        & \multicolumn{1}{r|}{0.67}                                                    & \multicolumn{1}{r|}{\textbf{1.00}} & \multicolumn{1}{r|}{\textbf{2.48}}         & \multicolumn{1}{r|}{1.00}          & \multicolumn{1}{r|}{2.52}                  & \multicolumn{1}{r|}{1.00}          & \multicolumn{1}{r|}{2.57}                  & \multicolumn{1}{r|}{0.48}          & \multicolumn{1}{r|}{3.57}                  & \multicolumn{1}{r|}{\textbf{1.00}} & \multicolumn{1}{r|}{\textbf{2.48}}         \\ \cline{2-14} 
\multicolumn{1}{|c|}{}                        & Mod\_4                                     & \multicolumn{1}{l|}{Grid}                        & \multicolumn{1}{r|}{2.88}                                                     & \multicolumn{1}{r|}{0.09}          & \multicolumn{1}{r|}{5.46}                  & \multicolumn{1}{r|}{\textbf{0.61}} & \multicolumn{1}{r|}{\textbf{5.46}}         & \multicolumn{1}{r|}{0.45}          & \multicolumn{1}{r|}{6.77}                  & \multicolumn{1}{r|}{0.30}          & \multicolumn{1}{r|}{7.88}                  & \multicolumn{1}{r|}{0.09}          & \multicolumn{1}{r|}{5.46}                  \\ \cline{2-14} 
\multicolumn{1}{|c|}{}                        & Mod\_5                                     & \multicolumn{1}{l|}{Grid}                        & \multicolumn{1}{r|}{1.21}                                                & \multicolumn{1}{r|}{\textbf{1.00}} & \multicolumn{1}{r|}{\textbf{3.19}}         & \multicolumn{1}{r|}{1.00}          & \multicolumn{1}{r|}{3.30}                  & \multicolumn{1}{r|}{0.75}          & \multicolumn{1}{r|}{3.24}                  & \multicolumn{1}{r|}{0.33}          & \multicolumn{1}{r|}{6.12}                  & \multicolumn{1}{r|}{\textbf{1.00}} & \multicolumn{1}{r|}{\textbf{3.19}}         \\ \hline
\multicolumn{1}{|c|}{\multirow{5}{*}{Area}}   & Mod\_1                                     & \multicolumn{1}{l|}{Area}                        & \multicolumn{1}{r|}{0.77}                                                   & \multicolumn{1}{r|}{0.00}          & \multicolumn{1}{r|}{2.63}                  & \multicolumn{1}{r|}{0.00}          & \multicolumn{1}{r|}{2.73}                  & \multicolumn{1}{r|}{\textbf{0.58}} & \multicolumn{1}{r|}{\textbf{2.52}}         & \multicolumn{1}{r|}{0.43}          & \multicolumn{1}{r|}{6.43}                  & \multicolumn{1}{r|}{\textbf{0.58}} & \multicolumn{1}{r|}{\textbf{2.52}}         \\ \cline{2-14} 
\multicolumn{1}{|c|}{}                        & Mod\_2                                     & \multicolumn{1}{l|}{Area}                        & \multicolumn{1}{r|}{2.46 }                                                   & \multicolumn{1}{r|}{0.04}          & \multicolumn{1}{r|}{292.01}                & \multicolumn{1}{r|}{0.04}          & \multicolumn{1}{r|}{300.02}                & \multicolumn{1}{r|}{\textbf{0.54}} & \multicolumn{1}{r|}{\textbf{300.02}}       & \multicolumn{1}{r|}{0.26}          & \multicolumn{1}{r|}{293.70}                & \multicolumn{1}{r|}{\textbf{0.54}} & \multicolumn{1}{r|}{\textbf{300.02}}       \\ \cline{2-14} 
\multicolumn{1}{|c|}{}                        & Mod\_3                                     & \multicolumn{1}{l|}{Column}                      & \multicolumn{1}{r|}{1.87}                                                     & \multicolumn{1}{r|}{0.01}          & \multicolumn{1}{r|}{4.44}                  & \multicolumn{1}{r|}{0.58}          & \multicolumn{1}{r|}{4.42}                  & \multicolumn{1}{r|}{\textbf{0.58}} & \multicolumn{1}{r|}{\textbf{3.66}}         & \multicolumn{1}{r|}{0.43}          & \multicolumn{1}{r|}{26.34}                 & \multicolumn{1}{r|}{0.58}          & \multicolumn{1}{r|}{3.66}                 \\ \cline{2-14} 
\multicolumn{1}{|c|}{}                        & Mod\_4                                     & \multicolumn{1}{l|}{Grid}                        & \multicolumn{1}{r|}{0.84}                                                  & \multicolumn{1}{r|}{\textbf{0.30}} & \multicolumn{1}{r|}{\textbf{4.19}}         & \multicolumn{1}{r|}{0.30}          & \multicolumn{1}{r|}{4.33}                  & \multicolumn{1}{r|}{0.30}          & \multicolumn{1}{r|}{4.35}                  & \multicolumn{1}{r|}{0.53}          & \multicolumn{1}{r|}{11.27}                 & \multicolumn{1}{r|}{\textbf{0.30}} & \multicolumn{1}{r|}{\textbf{4.35}}       \\ \cline{2-14} 
\multicolumn{1}{|c|}{}                        & Mod\_5                                     & \multicolumn{1}{l|}{Area}                        & \multicolumn{1}{r|}{0.97}                                                    & \multicolumn{1}{r|}{0.68}          & \multicolumn{1}{r|}{2.94}                  & \multicolumn{1}{r|}{\textbf{1.00}} & \multicolumn{1}{r|}{\textbf{2.88}}         & \multicolumn{1}{r|}{1.00}          & \multicolumn{1}{r|}{2.92}                  & \multicolumn{1}{r|}{0.34}          & \multicolumn{1}{r|}{3.53}                  & \multicolumn{1}{r|}{1.00}          & \multicolumn{1}{r|}{2.92}              \\ \hline
\end{tabular}
}

\caption{ Experimental results of 21 invoices classified by best graph-based approach.}
\label{tab:results}
\end{table*}

As expected, Table \ref{tab:results} shows that Dynamic strategy obtains really interesting results in terms of table detection (this aspect has of course been checked manually but real output examples cannot been shown here). Best results are obtained for 13 instances coming from all the family. Of course, results obtained for the strategy corresponding to the family are always the same as those for the Dynamic approach.
The surrogate function (OpenCV+Deduced Structures) works very well except for the Empty family where some characteristics are found and then induce a wrong strategy. Score for Dynamic strategy on the Empty family is 0 when the optimal strategy is not found by the surrogate function. An explanation is certainly the fact that the surrogate function returns a wrong structure where no table is present. The running time for the surrogate approximation is quite stable (between 0 to 3 seconds). However, the running time for the graph-based approach range from $2.31$ to $300$ seconds. This is due to the size of the structures returned by the surrogate function.

\section{Conclusion}

We propose a two-step approach for the extraction of tables in invoices using two complementary techniques. The table can indeed be detected at two levels : either by detecting some visual characteristics or by detecting the structural organization of the tokens of information, aggregated according to rows and columns that may intersect. Hence we have proposed a complete formalization of the possible relations between tokens in the document, once they have been extracted by classic document processing techniques. Our model allows us to benefit from an efficient graph-based representation where tables correspond to particular patterns of sub-graphs. Our method, which uses a visual surrogate estimation of the document to help a more structural combinatorial search of a table, shows good results on a set of representative difficult invoices. Future work includes finer visual surrogate evaluation and a better use of expert knowledge to characterize the elements of tables.


\end{document}